\documentstyle[11pt]{article}
\newcommand{\bea}{\begin{array}{c}}
\newcommand{\eea}{\end{array}}
\newcommand{\beq}{\begin{equation}}
\newcommand{\eeq}{\end{equation}}
\newcommand{\beqa}{\begin{eqnarray}}
\newcommand{\eeqa}{\end{eqnarray}}

\newcommand{\eq}{{equation~}}

\newcommand{\lexp}{\mathop{\bigl\langle}}
\newcommand{\rexp}{\mathop{\bigr\rangle}}
\newcommand{\gtilde}
	{\mathrel{\raisebox{-1ex}{$\stackrel{\textstyle >}{\sim}$}}}

\newcommand{\nbar}{\bar{n}}
\newcommand{\Nbar}{\bar{N}}
\newcommand{\xibar}{\bar{\xi}}

\def\Mpc{{\,h^{-1}\,{\rm Mpc}}}
\def\etal{{\it et al.\ }}

\def\lim{{\rm lim}}
\def\min{{\rm min}}
\def\max{{\rm max}}
\def\tot{{\rm tot}}

\def\h{{\,\rm h}}
\def\km{{\,\rm km}}
\def\Jy{{\,\rm Jy}}
\def\Mpc{{\,h^{-1}\,{\rm Mpc}}}
\def\s{{\,\rm s}}
\def\sr{{\,\rm sr}}

\font\eBF=cmmib10 scaled 1100
\newcommand{\r}{\hbox{\eBF r}}

\begin{document}
\baselineskip 18pt
\begin{titlepage}

\begin{flushright}
FERMILAB-Pub-93-097-A\\
May 1993\\
\end{flushright}

\smallskip

\begin{center}
{\Large\bf Redshift distortions of galaxy correlation functions}
\end{center}

\begin{center}
{\large\sc J. N. Fry%
\footnote{${}_{\strut}$ NASA/Fermilab Astrophysics Center,
Fermi National Accelerator Laboratory, Batavia, IL 60510-0500, USA}%
${}^,$\footnote{${}_{\strut}$ Department of Physics, University of %
Florida, Gainesville, FL 32611, USA}
and Enrique Gazta\~naga%
${}^{1,}$\footnote{${}_{\strut}$ Department of Physics, %
University of Oxford, Keble Road, Oxford OX1 3RH, England, UK}}
\end{center}

\medskip

\begin{center}
{\small\rm ABSTRACT}
\end{center}

\smallskip

To examine how peculiar velocities can affect the 2-, 3-, and 4-point
redshift correlation functions, we evaluate volume-average correlations
for configurations that emphasize and minimize redshift distortions for four
different volume-limited samples from each of the CfA, SSRS, and IRAS
redshift catalogs. We present the results as the correlation length
$ r_0 $ and power index $ \gamma $ of the 2-point correlation,
$\xibar_2 = (r_0/r)^{\gamma} $, and as the hierarchical amplitudes of
the 3- and 4-point functions, $ S_3 = \xibar_3/\xibar_2^{\,2} $ and
$ S_4 = \xibar_4/\xibar_2^{\,3} $.

We find a characteristic distortion for $\xibar_2$: the slope
$\gamma$ is flatter and the correlation length is larger in redshift
space than in real space; that is, redshift distortions ``move''
correlations from small to large scales.
At the largest scales (up to $12 \Mpc$), the extra power in the redshift
distribution is compatible with $\Omega^{4/7}/b \approx 1$. We estimate
$\Omega^{4/7}/b$ to be $0.53 \pm 0.15$, $1.10 \pm 0.16$ and $0.84 \pm
0.45$ for the CfA, SSRS and IRAS catalogs.

Higher order correlations $\xibar_3$ and $\xibar_4$ suffer similar
redshift distortions, but in such a way that,
within the accuracy of our analysis,
the normalized amplitudes $S_3$ and $S_4$ are insensitive to this effect.
The hierarchical amplitudes $S_3$ and $S_4$ are constant as a function
of scale between 1--$12 \Mpc$ and have similar values in all samples and
catalogues, $S_3 \approx 2$ and $S_4 \approx 6$,
despite the fact that $\xibar_2$, $\xibar_3$, and $\xibar_4$ differ
from one sample to another by large factors (up to a factor of 4 in
$\xibar_2$, 8 for $\xibar_3$, and 12 for $\xibar_4$).

The agreement between the independent estimations of $S_3$ and $S_4$
is remarkable given the different criteria in the selection of
galaxies and also the difference in the resulting range of densities,
luminosities and locations between samples.

\bigskip

\noindent{\it Subject Headings}:
Large-scale structure of the universe --- galaxies: clustering

\end{titlepage}

\newpage
\baselineskip 18pt

\section{Introduction}

Measurements of the galaxy 2-, 3-, and 4-point correlation functions,
$\xibar_2$, $\xibar_3$, and $\xibar_4$, indicate that a hierarchical
clustering structure, with $ \xibar_J = S_J \, \xibar_2^{J-1} $, holds
both in angular catalogs (Groth \& Peebles 1977; Fry \& Peebles 1978;
Szapudi, Boschan, \& Szalay 1992; Meiksin, Szapudi, \& Szalay 1992)
and in redshift catalogs (Bouchet, Davis, \& Strauss 1993; Gazta\~naga
1992), and both for IRAS and optically selected galaxies.
The values obtained for the hierarchical amplitudes $S_J$ do not
completely agree between different analyses (see Table 8 below),
and the question arises whether this is caused by redshift
distortions or by differences in the techniques and selections employed,
or arises from intrinsic differences in the samples.

Previous analyses of galaxy correlation functions have focused either
on the projected distribution inferred from angular data or on the
redshift distribution, but there are very few cases where both angular
and redshift correlations have been estimated for the same sample
(one notable exception for the 2-point correlation function is
Davis \& Peebles 1983).
Deriving statistical properties from angular data requires many
galaxies to compensate for the decrease in the projected amplitude,
and introduces uncertainties from the selection function.
With the addition of redshift information, the selection function is
better known, but the total number of galaxies in redshift samples is
usually orders of magnitude smaller because of the larger observational
time required to take a single redshift.
As a result, it is difficult to perform both analyses for a single catalog.
In particular, for the 3- and 4-point correlation functions we do not
know of any previous reference where both redshift and angular analyses
are considered over identical data.

Statistics of galaxy clustering inferred from redshift data are
likely to be distorted radially by peculiar velocities.
One known effect is the ``fingers of God'': the large velocities
in a tightly bound cluster disperse the apparent positions of galaxies
along the line of sight, such that in the extreme, clusters appear as
narrow needles pointed back at the observer.
Since typical peculiar velocities are a few hundred kilometers per second,
one might hope that by avoiding clusters or by looking on larger scales
this effect could be rendered unimportant.
However, fluctuations are also smaller on large scales and there are
coherent streaming effects.
Kaiser (1987) showed in perturbation theory that on large scales
velocity distortions are of the same order as density fluctuations
and distort large scale perturbation amplitudes by a factor
$ [1 + f(\Omega)\cos^2 \theta]$, where $ \theta $ is the angle
between the perturbation wave vector and the line of sight and
$ f(\Omega) \approx \Omega^{4/7} $.
Averaged over a spherical volume, fluctuations in redshift space
are thus enhanced by a factor $ [1+f(\Omega)/3] $.
Thus, we expect possible redshift distortions on all scales.

The effect on higher order moments is less clear.
Naively, the Kaiser analysis would seem to indicate that the $J$-point
hierarchical redshift amplitudes $S_J^z$ should be smaller than the real
amplitudes $S_J$, by a factor $ S_J^{\,z}/S_J \sim  [1+f(\Omega)/3]^{1-J}$,
and therefore be sensitive to $\Omega$.
However, extending the analysis to second order perturbation theory,
required to compute the 3-point function consistently, Bouchet \etal
(1992) claim that the net effect of redshift distortions on $S_3$ is
small.
Numerically, Lahav \etal (1993) find in simulations on smaller scales
that $S_3$ is quite distorted in redshift space, while Coles \etal (1993)
do not.
Further complicating the theoretical expectations is the possibility
of biasing, i.e. whether or not galaxies faithfully trace the matter
distribution.
For large scales, Fry and Gazta\~naga (1993) have shown that a local
biasing preserves the hierarchical structure to all orders but can
change the values of the $S_J$.
Kaiser's effect, arising from the peculiar velocity modulation of the
distribution, applies equally to the galaxy number distribution, even
when galaxies are not unbiased tracers of the mass, and in general all
these effects are mixed at each order.

To separate redshift distortions of observed correlations from other
sources of difference, we believe that it is important to perform a
dual analysis to the same data samples.
Thus, in this paper we present an analysis in which identical techniques
of estimation are applied over the same data samples to compute
volume average correlations in redshift space for volumes that
are expected to exhibit redshift distortions, and for configurations
that are expected to minimal distortions.
In \S~2 we discuss measuring correlations from moments of
counts in cells, including an analytic extension of the count-in-cell
tails for finite volume samples.
In \S~3, we relate volume average parameters to intrinsic parameters for
spherical and conical cells for power-law underlying correlations.
Spherical cells are expected to experience redshift distortions,
which should be minimal in conical cells.
\S~4 presents details of the samples we use and the data analysis.
\S~5 contains a summary of results and a final discussion.

\section{Estimation of volume-average correlations}

We center our analysis on the the volume-average correlation functions,
\beq
\xibar_J(V) = {1\over{V^J}} \int_V d^3r_1 \dots d^3r_J \,
\xi_J(\r_1,\dots ,\r_J) .  \label{xibarj}
\eeq
We use volume-limited samples so that no selection
function appears.
We assume that every region in redshift space is equally weighted
and faithfully traced by the galaxies in each sample.

We use the count-in-cell probabilities $P_n(V)$ to compute moments
and obtain the $\xibar_J(V)$.
The correlation functions $ \xibar_J(V) $ can be found from
the count-in-cell probabilities $P_n(V)$, the probability to find
$n$ galaxies in a randomly selected cell of volume $V$,
from moments $ \lexp (\Delta n)^J \rexp $.
Including discreteness contributions (cf. Peebles 1980; Fry 1985),
the first few moments are related to the $\xibar_J(V)$ as
\beqa
\lexp n \rexp  &=& \Nbar \nonumber \\*
\lexp (\Delta n)^2\rexp  &=& \Nbar^2\xibar_2+\Nbar \nonumber  \\*
\lexp (\Delta n)^3\rexp  &=& \Nbar^3\xibar_3+3\Nbar^2\xibar_2+\Nbar
\nonumber \\*
\lexp (\Delta n)^4\rexp - 3 \lexp (\Delta n)^2\rexp {}^2
&=& \Nbar^4\xibar_4 + 6\Nbar^3\xibar_3 + 7\Nbar^2\xibar_2 + \Nbar
\label{eq:moments}
\eeqa

\subsection{Count-in-cell probabilities}

We estimate $P_n(V)$ in the following way: given the radius
of a test-sphere, we choose randomly a center inside the survey sample,
count the number of galaxies found inside the cell, and accumulate the
number of cells $N_n$ with $ n = 0 $, 1, 2, $ 3 \dots $ galaxies.
We repeat this procedure $N_T$ times, to be chosen so that the
sampling spheres will overlap with each other.
Counts-in-cell probabilities are then estimated by $ P_n = N_n/N_T $.
The precision of this estimation is limited by the number of
{\it independent} cells; from the Poisson distribution, we expect
$ \Delta P_n /P_n \gtilde (P_n N_T)^{-1/2}$ for $P_n \ll 1$.
We have to choose $N_T$ in order to estimate counts in cells with the
right accuracy.
A better choice is $ N_T = \xibar_2^{\,-3/2} V_T/V $
(Gazta\~naga \& Yokoyama 1993), but because $\xibar_2$ is not known a
priori we must iterate the process to obtain $N_T$, using as a first
approximation the value of $\xibar_2$ estimated from pair-counts.
The results do not differ much between these two prescriptions, and
therefore we have chosen the one that is more simple, i.e. $N_T=2 V_T/V$.
Error bars for the $P_n$ are estimated from 90\% confidence in different
realizations of the positions of the independent cells,
which agree well with the binomial estimate of errors above.

\bigskip

\subsection{Estimation of moments}

Once we have the $P_n(V)$, we next calculate moments of different
orders to estimate the correlation functions from \eq(\ref{eq:moments}).
This brings us to the finite sample problem.
The number of independent cells $N_T$ used to estimate $P_n(V)$ is
necessarily finite, because the size of the sample is finite.
This limits the smallest value of $P_n$ that can be estimated,
$ P_\min (V_c) \sim 1/N_T $.
There will also be a maximum cell count, $ n_\max $.
The probabilities $ P_n $ for counts $ n > n_\max $ and with
$P_n < P_\min $ can not be estimated, and thus the moments,
nominally given by
\beq
\lexp n^m \rexp = \sum_{n=0}^\infty n^m P_n \, ,
\label{eq:moment}
\eeq
can only be summed up to $ n = n_\max $.
An estimate of when this is an important effect is whether
$ n_\max $ is small relative to $ N_c = \Nbar\xibar_2 $,
the number of galaxies in a ``typical'' cluster.
If $ n_\max < N_c $, then the moments will be systematically
underestimated, by an increasing amount for higher moments.
When only a few rich clusters such as Virgo or Coma are present in
a catalog (such as the CfA) there may also be a ``bump'' in the
tail of $P_n$.
This ``bump'' may introduce a large artificial enhancement of the
correlation functions.
A further related problem is interaction with the boundary:
when cell locations are restricted to avoid intersection with the
boundaries, the size of the survey area covered gets smaller as the
size of the cell is increased.
This can introduce additional bias, as the center area of the
sample is more weighted than the boundary, and, moreover, reduces
even further the number of independent cells, so that
$ P_\min $ is larger and $ n_\max $ is smaller,
which again eventually results in underestimated moments.
When the sample is large enough that finite sample size is not a problem,
the $P_n$ appear to fall exponentially with $n$.
This suggests a method to correct for the limitations of finite volume.

Modeling the tail, that is, extrapolating the $P_n$ to $n > n_\max$,
can help to compensate for the underestimation, or, at the least,
can be used to study systematic errors in the moments.
Both observations and various scale-invariant models suggest that
the probability tails are exponential, behaving as
$ P_n \propto e^{-n/ N_c} $, where $ N_c = \Nbar\xibar_2 $.
Thus, when $ n_\max > N_c  $, the contribution to $ \sum n^m P_n $
from $ n_\max $ to $ \infty $ is small when compared with the whole
sum in \eq(\ref{eq:moment}), as claimed above.
Modeling also reduces the effects on the correlations that appears
when rich clusters such as Coma are present.
By fitting an exponential to the tail, the effect of a ``bump''
in the $ P_n $ is diluted.

We fit the tail of the observed counts-in-cells to an exponential
function, $ \lexp N_n \rexp = \mu_n = C e^{-\alpha n}$,
using a Poisson likelihood function,
\beq
{\cal L}= 
\sum_n \log\left[ {1 \over{N_n!}} \mu_n^{N_n}e^{-\mu_n} \right],
\eeq
where $N_n$ are the observed counts (the number of cells with
$n$ galaxies) and $ \mu_n = C e^{-\alpha n} $.
The values of $C$ and $\alpha$ are found by maximization of
the likelihood $ {\cal L} $.
As a test we can verify that the observed counts in the tail follow a
Poisson distribution around the exponential.
In practice we construct a routine to find the value $ n_1 $ where the
``tail'' begins, i.e. the value of $i$ where the counts begin to fit an
exponential (ideally, $ \Nbar \ll n_1 \ll n_\max $),
fit $ \mu_n = C e^{-\alpha n} $ for $ n_1 < n <  n_\max $,
and extrapolate $ P_n \sim \mu_n $ for $ n > n_\max $.
We then must renormalize the distribution of counts so that
$ \sum_0^\infty \bar{P_n} = 1 $.
For small cell volume, the resulting correction to the moments is very
small, as $n_\max$ is already quite large, while for large cells, both the
contribution from and the uncertainties in this modeling could be very large.
To estimate the systematic errors introduced in the moments by modeling
the exponential tail, we report in the results the difference between the
``raw'' moments (without modeling of the tails) and the modeled moments,
\beq
\Delta \lexp n^m \rexp=\left| \sum_{n=0}^\infty n^m \bar{P_n}
- \sum_{n=0}^{n_\max} n^m P_n  \right| .  \label{eq:errorm}
\eeq
This provides a natural way to measure when the finite volume
effects are large.

\section{Volume average correlations}

\subsection{Redshift vs. real space correlations}

The choice of volume shape can emphasize or minimize the effects of
peculiar velocities.
As one choice, we compute moments over spheres of radius
$R$ in redshift space.
As remarked above, the contribution of peculiar velocities to the apparent
redshift distance are expected to distort the resulting statistics.
We refer to the results from spherical volumes as ``redshift'' results.
To extract information about correlations from a redshift catalog that
is not distorted by the peculiar velocity modulation of the distribution,
we simply change the shape of the volume, to cells with the shape of
conic sectors of a sphere between radii $ d < r < D $, of opening angle
$\theta$ around a radial axis (in our analysis, $d$ is a fixed to
$ 2500 \km \s^{-1} $ and $D$ is the depth of the sample; see Table~1).
The estimated counts-in-cells for these volumes are not distorted
except perhaps at the inner and outer radial boundaries of the cell,
where peculiar velocities can move a few galaxies in or out of the cell,
a small effect for deep samples.
Because results from these conic volumes are not distorted by peculiar
velocities, we refer to them in brief as ``real space'' results.
As will become apparent below, the conic cell analysis is similar to
considering just the projected, magnitude-limited angular distribution,
with integrals similar to those in the standard angular analysis.
However, since redshift information is used to determine the distance
and absolute magnitudes of the galaxies in our samples,
the resulting analog of the Limber equation (see Peebles 1980) for these
volume-limited samples does not involve an integral over the
luminosity function, thus removing a source of uncertainty.

\subsection{Power-law correlations}

In extracting the underlying correlation functions from the
volume-averaged results, we assume a power-law two-point correlation
function, $ \xi_2(r) = (r_0/r)^\gamma $, and a hierarchical three-point
function, as suggested by our data (see Figs.~$1a$--$c$).
The power law model $ \xi_2(r) = (r_0/r)^\gamma $ is not a good
approximation for all possible ranges of scales
(see i.e. Groth \& Peebles 1977; Maddox \etal 1990) but, given the
resolution in our analysis, it is quite a good one for the range of scales
we have been able to inspect; i.e. $ R \simeq 1$--$ 20 \Mpc$.

The volume-average two-point correlation is
\beq
\xibar_2(V) = {1\over{V^2}} \int_V d^3r_1 d^3r_2 \, \xi_2(|\r_1 - \r_2|) .
\label{xibar2}
\eeq
For $ \xi_2(r) = (r_0/r)^\gamma $ and spherical cells of radius $R$,
we have
\beqa
\xibar_2^{S} &=& S(\gamma) \, (r_0/R)^\gamma \nonumber \\
S(\gamma) \!\!\! &=& {{2^{-\gamma}} \over
{(1-\gamma/3)(1-\gamma/4)(1-\gamma/6)}}, \label{eq:xibar2R}
\eeqa
so that $\xibar_2^{S}(R) $ is also a power law, with, in the absence of
peculiar velocity distortions, the same slope as $ \xi_2(r) $ and with
amplitude different by a factor $ S(\gamma) $ of order unity.
For $ \gamma = 2 $, $ S(\gamma) = 9/4 $;
for $ \gamma = 1.5 $, $ S \approx 1.5 $.

For a volume shaped as conic sector of an sphere of opening angle
$\theta$ around a radial axis between radii $ d < r < D $ we find for
power-law $ \xi_2(r) $ in the limit $ \theta \ll 1 $,
\beqa
\xibar_2^{C} &=& C(\gamma) \, \theta^{1-\gamma}  \nonumber \\
C(\gamma) \!\! &=& {I_1 \over I_0^2} \, {{H_\gamma}\over{2\pi}}  \,
\left({{r_0} \over D} \right)^\gamma \nonumber \\
H_\gamma \! &=& \int_{-\infty}^\infty {{dx} \over {(1 + x^2)^{\gamma/2}}} ~=~
{{\Gamma(1/2) \, \Gamma(\gamma/2-1/2)} \over {\Gamma(\gamma/2)}} \nonumber \\
I_k &=& {{1-(d/D)^{3(k+1)-k\gamma}} \over{3(k+1)-k\gamma}}
\int_0^1 \! z \, dz \, F^k(z) \nonumber \\
F(z) \!\! &=& \int_0^1 \! x \, dx \int_0^{2 \pi} d\phi \,
(z^2 + x^2 - 2 z x\cos\phi)^{(1-\gamma)/2}
\label{eq:xibar2t}
\eeqa
so that $ \xibar_2^{C} $ is also a power law but with a different
slope and (possibly greatly) reduced amplitude,
$ \xibar_2^C(\theta) \sim \xi(D) \, \theta^{1-\gamma} $.
In each case, from the fitted slope and amplitude of observed $\xibar_2$
and from the expressions above, one can estimate the values of $r_0$ and
$\gamma$ for the intrinsic two-point correlation function.

For higher order correlations we assume hierarchical properties,
$ \xibar_J = S_J \xibar_2^{J-1} $, again in agreement with our data
(see Figs.~2--3) and previous analyses.
For the 3-point correlation function we use the local hierarchical
expression
\beq
\xi_3(\r_1,\r_2,\r_3)= Q_3 \, \left[\xi_2(r_{12})\xi_2(r_{13}) +
\xi_2(r_{12})\xi_2(r_{23}) + \xi_2(r_{13})\xi_2(r_{23}) \right]
\label{Q}
\eeq
to compute the volume average
\beq
\xibar_3(V) = {1\over{V^3}} \int_V d^3r_1 d^3r_2 d^3r_3 \,
\xi_3(\r_1,\r_2,\r_3) . \label{xibar3}
\eeq
We are interested in the hierarchical amplitudes, i.e. we want to estimate:
$ S_3 = \xibar_3(V) / \xibar_2^{/,2} (V) $.
For a spherical volume, the hierarchical relation (\ref{Q})
with $\xi_2(r)=(r_0/r)^\gamma$ implies:
\beqa
S_3^{S} &=&  3 \, \eta(\gamma) \, Q_3  \nonumber \\
\eta(\gamma) \! &=& \frac{(6-\gamma)^2}{3(2-\gamma)^2}\left[
\frac{46-48\gamma+17\gamma^2-2\gamma^3}{2(7-2\gamma)(9-2\gamma)}\right. \\
&& \qquad -\left.\frac{2^{2\gamma} \sqrt{\pi}\,(2-\gamma) \Gamma(5 -\gamma)}
{256 \, \Gamma(9/2-\gamma)} -\frac{2^{2\gamma}\sqrt{\pi}\,\Gamma(5-\gamma)}
{256 \, \Gamma(11/2-\gamma)} \right] \nonumber \label{eq:xibar3R}
\eeqa
(Gazta\~naga \& Yokoyama 1993).
For $0 \le \gamma \le 2 $, $S_3$ and $3 Q_3$ differ by less than 3\%,
so that in practice one can use as a good approximation
$S_3^{S} \simeq 3 Q_3 $.
In our analysis, averages over spherical volumes and thus $S_3^S$ are
obtained in redshift space.

For the conic cell shape we have, in the limit of small $\theta$,
\beqa
S_N^C &=& N^{N-2} Q_N
{ {I_0^{N-2} \, I_{N-1}}\over{I_1^{N-1}}  }
\label{eq:SNC}
\eeqa
where $I_N$ is defined in \eq (\ref{eq:xibar2t}) above.
Thus, by measuring $ S_3^{C} $ and $ \gamma $ for conic cells one can
use the above relations to estimate the corresponding $Q_3$ or $S_3^S$
in real space.

\section{Data analysis results}

\subsection{Catalogues}

We use three different catalogs of galaxies:

\begin{itemize}

\item $i$) the North Zwicky Center for Astrophysics catalog
(Huchra \etal 1983, hereafter CfA), with $m_B < 14.5$,
$\delta \ge 0$, and  $b^{\rm II} \ge 40^\circ $, which has a solid
angle of $ 1.83\sr $.

\item $ii$) the Southern Sky Redshift Survey (Da Costa \etal 1991, hereafter
SSRS) of diameter-selected galaxies with $ \log d(0) \geq 0.1 $ from the
ESO catalog with $\delta \le -17.5^\circ$, $b^{\rm II} \le -30^\circ $;
the solid angle is $ 1.75 \sr $.

\item $iii$) the full-sky coverage redshift survey of galaxies from the
IRAS ({\it Infrared Astronomical Satellite}) database
(Strauss \etal 1992, hereafter IRAS) with limiting flux density of
$ 1.936 \Jy $ at $ 60 \, \mu$m and galactic latitude $ |b| >5^\circ$;
the solid angle is $ 11.0 \sr $.

\end{itemize}

We select four different volume-limited subsamples from each catalog,
as shown in Tables 1--3, where $ z_\lim $ is the maximum redshift,
$N_\tot $ is the total number of galaxies, $V_T$ is the volume of each
sample, and $M_B$, $d_\lim$, and $z^2 f_{60}$ are the limiting absolute
magnitude, ``face-on'' diameter, and flux, respectively.
Lengths in the tables quoted as ``Mpc'' use $ H_0 = 100 $.
To produce fair volume-limited samples we have not included the fainter or
smaller galaxies, so that the space in each sample is homogeneously filled.
For example, each CfA sample includes galaxies brighter than
$M_B = m_B - 25 - 5\log(z_\lim/H_0) $, where $m_B = 14.5$ is the limiting
apparent magnitude.
In addition, galaxies in the CfAN50 sample are restricted to be fainter
than $ M_B = -20.3 $, so that this sample is independent of CfAN90.
In the SSRS catalogue, each sample includes galaxies with physical diameter
greater than $ d_\lim = z_\lim \theta_{\rm cut} / H_0 $,where
$\theta_{\rm cut} = 1 {'.} 26 $ is the ``face-on'' diameter cut-off.
In the IRAS  catalogue, each sample includes galaxies with absolute flux
$ z^2 f_{60} $ greater than the limiting flux $ z^2_\lim \, 1.936 \Jy $.

Heliocentric redshifts are corrected only for our motion with respect to
the rest frame of the Cosmic Microwave Background;
$ v = 365 \km \s^{-1} $ toward $ (\alpha,\delta) = (11.2^\h, -7^\circ) $
(Smoot \etal 1991).
Galaxies with redshifts smaller than $ 2500 \km\s^{-1} $ are excluded,
since at these scales typical peculiar velocity flows (of several hundreds
of km/s) can compete with the general recession velocity.

\subsection{Count-in-cell volumes}

For each sample, we want to estimate the count-in-cell probabilities
$P_n$ as a function of the cell volume.
The total number of degrees of freedom is limited by the density $\nbar$
of the sample and its total volume $V_T$, so that the size of the cell is
bounded bellow by $\nbar$ and above by $V_T$.
The number of different cell sizes and their intervals should not be set
arbitrarily, as the resulting moments or correlations at each point would
not be independent. For samples of small density,
a good  criterion for statistically independent sizes is
$ V_i = i / \nbar $, for $ i = 1,2,\dots $, as then each new volume
includes on average one more galaxy.
Notice how different this is from using equal spacing in the radius
of spherical cells for all samples; this not only introduces a
different weighting but also imposes  an arbitrary number of degrees
of freedom, regardless of the size or density of the sample.
In our analysis we use equal volume spacing in units of $ V_1 = 1/\nbar $.
We use half this unit to study the fluctuations
or when we have only few points.
The largest cell to be used is limited by the boundaries of the sample.
Using different boundary conditions,  Gazta\~naga (1992) and
Gazta\~naga \& Yokoyama (1993) have found that a good phenomenological
bound is $ V \leq V_T/3^3 $, so that the diameter of the cell is at most
one third of the typical size of the sample.
This is just one upper limit; we also discard the larger cells of our
analysis when the estimated uncertainties, from \eq (\ref{eq:errorm})
are too large.
The resulting range of scales displayed in Tables 4--6 is different
for each sample.
The larger samples can accommodate larger cells, but also have smaller
densities and thus do not have information on the smaller scales.

\subsection{Correlation functions}

We estimate the correlation functions using the moments of counts-in-cells
as explained in section \S~2.
Figures $1a$--$c$ show the volume average two-point correlation
function, $\xibar_2$, as a function of the size of the cell.
Open squares correspond to spherical (``redshift'') cells and
triangles correspond to conic (``real space'') cells.
The scale of the cell, given in Mpc with $ H_0 = 100 $, corresponds to
the radius $R$ of the redshift-space sphere or the radius of the base
of the conical sector, $ \theta D $ with $ D $ the depth of the sample.
The error of each point is the maximum between the systematic error
(\ref{eq:errorm}) from correcting the count distribution tail for finite
volume and the statistical error from the $P_n$ propagated to the moments.
The dashed line in each figure is the best fit to a power law,
weighted by the errors.
The agreement with a power law model justifies our assumption
of the relations in section \S~3.
The slopes for the conic cells (triangles) and spherical shells
(squares) follow what we expect from the projection,
$\xibar_2^C \sim \theta^{1-\gamma} D^{-\gamma}$ and
$\xibar_2^S \sim R^{-\gamma}$.
In Tables 4--6 we display the values of the fit in terms of the slope
$\gamma$ and correlation length $r_0$ of the intrinsic correlation function
$ \xi_2(r)=(r_0/r)^\gamma $ using \eq(\ref{eq:xibar2R}) and
\eq(\ref{eq:xibar2t}).
Note that the range of scales where the fit is done is different for
each sample, as explained above.
We quote both the statistic and the systematic error
(the last in parenthesis).
The errors in the fitted parameters, $r_0$ and $\gamma$, reflect both
the goodness of the fit and the statistic (or systematic) error in the
estimated correlations;
both contributions are usually of comparable magnitude.

Figures $2a$--$c$ show  $\xibar_3(R) $ (triangles) and $\xibar_4(R)$
(squares) for spherical (redshift) cells, plotted against $\xibar_2(R)$
for all samples.
In Figures $3a$--$c$ we present a similar plot for conical
(real space) cells.
Again, the error plotted at  each point is the maximum between the
systematic error (eq.~[\ref{eq:errorm}]) and the statistical error
from the $P_n$ propagated to the moments.
The dashed line is the best fit to $\xibar_3 = S_3 \, \xibar_2^2$
and $\xibar_4 = S_4 \, \xibar_2^3 $ weighted by the errors.
The agreement of these power law models is evidence in favor of the
hierarchical model and, again, justifies our assumptions in section
\S~3.

Tables 4--6 shows the parameters $S_3$ and $S_4$ of the fit.
The values of $S_3$ for conical cells have been scaled to the
corresponding value of a spherical cell using \eq(11)
and \eq(\ref{eq:xibar3R}), so that real space and redshift space
amplitudes can be directly compared.
The correction to convert $ S_3^{C} $ into $ S_3^{S} $ is smaller
than 15\% so that we have not propagated the uncertainties from
$ \gamma $ in this conversion.
Because this correction turns out to be small and there are large
uncertainties in $S_4$ we have not tried to calculate the exact
correction in the case of the 4-point function, but have just scaled
the correction found for $S_3$ using the dimensional argument
$S_4^{S}/S_4^{C} \simeq [S_3^{S}/S_3^{C}]^2 $.
We quote both the statistic and the systematic error
(the last in parenthesis).
The errors in the fitted parameters $S_3$ and $S_4$ reflect both the
goodness of the fit and the statistic (systematic) error in the
estimated correlations;
both contributions are usually of comparable magnitude.
Samples IRAS65 and IRAS80 have a very noisy signal for spherical cells
and results are not quoted.

The average values reported in Tables 4--6 are the mean values for
all samples within a catalog, weighted by $ 1 / \sigma_i^2 $
(using the maximum between the statistic and systematic errors).
The first error in the average is the weighted dispersion about the mean,
scaled to a 1-$\sigma$ interval after accounting for the difference
between the Gaussian and Student-$t$ distribution.
For most averages, with four degrees of freedom, the dispersion
is increased by the factor 1.14163.
In parentheses we also quote the expected error,
$ 1 / \sum (1/\sigma_i^2) $.
These two are the same when (reduced) $ \chi^2 = 1 $.

\subsection{Redshift distortions }

In the linear regime Kaiser (1987) has derived the expected ratio
of two-point function in redshift space, $\xi_2(s)$, to that in
real space, $\xi_2(r)$:
\beq
{\xi_2(s)\over{\xi_2(r)}}= 1 + {2\over{3}}{\Omega^{4/7}\over{b}}
+{1\over{5}}{\Omega^{8/7}\over{b^2}}, \label{eq:omega}
\eeq
where $b$ is the bias factor between matter fluctuations
and galaxy fluctuations, $\delta_g = b \delta_m$.
Lilje \& Efstathiou (1989) have shown that these redshift distortions
are expected even on mildly non-linear scales where
$ \xi_2 (r) \sim 1 $.
Non-linear effects and the ``fingers of God'' for smaller scales tend
to suppress the redshift correlations, so that $ \xi_2(s)/\xi_2(r)$
becomes smaller than 1 for small separations (Davis \& Peebles 1983).
Therefore, even if we have not reached the scales where the linear theory
should apply (whatever this scale is), we should be able to place a
lower bound on $\Omega^{4/7}/b$ from the maximum redshift distortion
we are able to detect.

As will become apparent, we can not use our data to make an accurate
determination of this effect, but we can estimate the order of magnitude.
At some point the power laws we have observed for the real and
redshift two-point correlations will break to reproduce
Kaiser's results, a constant ratio of $\xi_2(s)$ to $\xi_2(r)$.
The ratio we see on larger scales is not constant, but rising with
scale, except in one case.
Even though in our analysis this ration has not reached a constant
value, presumably because of poor sensitivity at large scales, we
can use the extra power observed in the redshift correlation to place a
lower bound on the factor $\Omega^{4/7}/b$, and, if $ \Omega = 1 $,
an upper bound on $b$.
Using the values of $r_0$, $s_0$ and $\gamma$ shown in Tables 4--6,
we estimate the ratio $ \xi_2(s) / \xi_2(r) $ at $ r_0 $ and at the
scale $r_{\max}$ corresponding to the largest spherical (redshift)
cell in each sample.
Table~7 shows $ r_\max $, the value of the intrinsic two-point
(real-space)  correlation, the correlation strength at that scale,
$ \xi_2(r_\max) $, and the resulting values of $\Omega^{4/7}/b$ resulting
from \eq(\ref{eq:omega}).
As argued above, the ratio $ \xi_2(s)/\xi_2(r) $ increases towards
larger scales, and thus the final column represents our strongest
lower bound on $ \Omega^{4/7}/b $.
Errors in $\Omega^{4/7}/b$ are propagated from the errors in
$r_0$, $s_0$, and $\gamma$ (again, the maximum between the statistic
and systematic errors).
The average values in Table~7 are the mean values for all samples
within a catalog, weighted by the errors.
The negative values of $ \Omega^{4/7}/b $ for CfAN65 do not affect the
average value much, but do increase the estimated dispersion by a factor
of 2.

\section{Discussion}

In this work, we have computed volume-averaged correlation functions
$\xibar_2$, $\xibar_3$ and $\xibar_4$ for conical and spherical
configurations that allow us to observe the effects of peculiar
velocity distortions.
We have considered volume-limited samples so that no assumption has to
be made about a luminosity function.
We make use all the degrees of freedom in each sample and weight
them properly.
To account for boundary and finite size effects we have introduced a
modeling of the tail of the distribution of fluctuations which helps
to improve the signal and estimate the uncertainties.
We report the difference between results obtained with and without this
modeling of the tail as a systematic error, which we have taken as the
minimum uncertainty in our results, a conservative approach that leaves
our conclusions safely unaffected by our assumptions.
To simplify comparisons of the analyses  of different samples we have
reduced the observed correlation functions, $\xibar_2(R)$, $\xibar_3(R)$,
and $\xibar_4(R)$ in Figures 1--3, to a few parameters shown in
Tables 4--6.

Our analysis reproduces the already  known trend (Davis \etal 1988)
that the amplitude $ s_0 $ of the two-point redshift correlation function
of optical galaxies grows with the depth of the survey, at least up to
a redshift of $ 8000 \km \s^{-1} $.
This trend is confirmed here for both the Southern and Northern
galactic hemispheres for optical selected galaxies, i.e. in both the SSRS
and CfA catalogs; within our poor resolution, it is not observed in
the IRAS samples.
As a new result, we find that this scaling with depth seems to occur also
in real space, which indicates that it can not just be a consequence of
redshift distortions.
We also find that for depths larger than $ 8000 \km \s^{-1} $ the
trend is broken: the observed correlation length decreases in the
$ 9000 \km \s^{-1} $ samples.
This indicates that the effect may be accidental or caused by sampling
an ``unfair'' local volume, but does not imply a simple ``fractal'' or
inhomogeneous behavior.

For the average values and even for separate subsamples within each
catalog, the intrinsic 2-point correlation length $ r_0 $ of the CfA and
SSRS are very similar.
The relative amplitude of the IRAS and CfA two-point correlations is in
very good agreement with other analyses (Strauss \etal 1992; Saunders,
Rowan-Robinson, \& Lawrence 1992) although the techniques and samples
used are quite different.
There is a significant difference between the average slope $\gamma$ of
$ \xi_2 \sim r^{-\gamma} $ in real space between the SSRS and the CfA
or IRAS catalogues.
The slope in the SSRS, $\gamma = 1.62 \pm 0.04 $, is lower than the
``standard'' value of $ \gamma=1.8 $, but the errors are quite large.
SSRS diameter magnitudes are more subject to plate-to-plate correction
which might introduce some large scale power.
There is also a very large void in this catalog
(larger than any in the CfA1).

In general, there is a characteristic redshift distortion on $\xi_2(r)$:
the slope $\gamma$ is flatter and the correlation length is larger in
redshift space than in real space, that is, redshift distortions ``move''
correlations from small to large scales.
In particular, for the average values, going from real to redshift space
reduces the slope of $\xibar_2$ by 30\% (SSRS), 20\% (IRAS) and 13\% (CfA).
At the largest scales the extra power in the redshift distribution
is compatible with a $ \Omega^{4/7}/b \simeq 1 $, as seen in Table~7.
As expected, the average values of $ \Omega^{4/7}/b $ for each catalog
are correlated with the amount of reduction in the redshift slope.
Our results for $ \Omega^{4/7}/ b $ in the CfA are much larger than
those deduced by Lilje \& Efstathiou (1989) from Figure~3 of
Davis \& Peebles (1983), perhaps because they only use scales up to
$ 7 \Mpc $, whereas ours reach to $ 12 \Mpc $.

The higher order correlations $\xibar_3$ and $\xibar_4$ are redshift
distorted also, but in such a way that the hierarchical amplitudes
$S_3$ and $S_4$ are insensitive to these distortions, within the
resolution of our analysis.
This is in agreement with the numerical studies of Coles \etal (1993).
It follows that redshift distortions are probably not responsible
for the discrepancies in the values of $S_3$ and $S_4$ found in some
of the previous analyses, summarized  in Table 8.
Moreover, the hierarchical amplitudes $S_3$ and $S_4$ are constant
as a function of scale between 1--$20 \Mpc$ (which includes the
transition between linear and non-linear scales), and they have a
similar value in all samples, even for different catalogues,
despite that $\xibar_2$, $\xibar_3$ and $\xibar_4$ differ
from one sample to another by large factors (up to factors of 4
in $\xibar_2$, or 8 for $\xibar_3$, or 12 for  $\xibar_4$).
This is remarkable given the different criteria in the selection of
galaxies and also the difference in the resulting range of densities,
luminosities and locations between samples.

\bigskip
\noindent
{\large\bf Acknowledgements}

\bigskip

After this work was completed we became aware of other work on exponential
modeling of the probability tail by Colombi, Bouchet, \& Schaeffer (1993),
who also test their prescription with numerical simulations.
E.~G. acknowledges useful discussions with Jun'ichi Yokoyama.
This work was supported in part by DOE and by NASA (grant NAGW-2381)
at Fermilab.

\newpage

\noindent
{\large\bf References}
\bigskip

\def\pp{\par\parshape 2 0truecm 17truecm 1truecm 16truecm\noindent}
\def\paper#1;#2;#3;#4; {\pp#1, {#2}, {#3}, #4}
\def\book#1;#2;#3;#4; {\pp#1, {\sl #2} (#3: #4)}
\def\preprint#1;#2; {\pp#1, #2.}

\pp Bouchet, F.~R., Davis, M., \& Strauss M. 1993,
in {\sl Proceedings DAEC Workshop}, ed. G. Mamon, in press
\paper Bouchet, F.~R., Juszkiewicz, R., Colombi, S., \& Pellat, R., 1992;%
ApJ;394;L5;
\pp Coles, P., Moscardini, L., Lucchin, F., Matarrese, S.,
\& Messina, A. 1993, preprint
\pp Colombi, S., Bouchet, F.~R., \& Schaeffer, R. 1993, A\&A (in press)
\paper Da Costa, L. N., Pellegrini, P., Davis, M., Meiksin, A.,
Sargent, W., \& Tonry, J. 1991;ApJ Suppl;75;935;
\paper Davis, M., \& Peebles, P.~J.~E. 1983;ApJ;267;465;
\paper Davis, M., Meiksin, A., Strauss, M. A., da Costa, L. N.,
\& Yahil, A. 1988;ApJ;333;L9;
\paper Fry, J.~N. and Peebles, P.~J.~E. 1978;ApJ;221;19;
\paper Fry, J.~N. 1985;ApJ;289;10;
\pp Fry, J. N., \& Gazta\~naga, E. 1993, ApJ (in press)
(FERMILAB-Pub-92/367-A)
\paper Gazta\~naga, E. 1992;ApJ;398;L17;
\paper Gazta\~naga, E., \& Yokoyama, J 1993;ApJ;403;450;
\paper Groth, E.~J., \& Peebles, P.~J.~E. 1977;ApJ;217;385;
\paper Huchra, J., Davis, M., Latham, D., \& Tonry, J. 1983;ApJS;%
52;89;
\paper Kaiser, N. 1987;MNRAS; 227;1;
\paper Lahav, O., Itoh, M, Inagaki, S., \& Suto Y. 1993;ApJ;402;387;
\paper Lilje, P.~B., \& Efstathiou, G. 1989;MNRAS;236;851;
\paper Maddox, S. J., Efstathiou, G., Sutherland, W.J. \&
Loveday, L. 1990;MNRAS;242;43p;
\paper Meiksin, A., Szapudi, I. \& Szalay, A. S. 1992;ApJ;394;87;
\pp Peebles, P.~J.~E. 1980, The Large-Scale Structure of the Universe
(Princeton Univ. Press)
\paper Saunders, W., Rowan-Robinson M., \& Lawrence A. 1992;MNRAS;258;134;
\paper Smoot, G.~F. {\sl et al.} 1991;ApJ;371;L1;
\paper Strauss, M.~A., Davis, M., Yahil, A. \& Huchra, J.~P.
1992;ApJ;385;421;
\paper Strauss, M.~A.,  Huchra, J.~P, Davis, M., Yahil, A., Fisher, K.,
\&Tonry, J. 1992;ApJS;83;29;
\paper Szapudi, I., Szalay, A.~S., \& Boschan, P. 1992;ApJ;
390;350;

\newpage

\def\df{\dotfill}
\def\hf{\hfill}

\begin{center}
TABLE 1\\
CfA Samples
\end{center}

\def\ph{{\phantom{.5}}}
\def\ss{\rule[0ex]{0mm}{3.5ex}\rule[-1.75ex]{0mm}{3.5ex}\\*}
\begin{center}
\begin{tabular}{|c c c c c|}
\hline
\makebox[7em]{Sample}&\makebox[6em]{$ z_\lim $ (km/s)}%
&\makebox[8em]{$M_B$} &\makebox[5em]{$N_{\rm gal}$}&%
\makebox[6em]{$V_T$ (Mpc$^3$)}\ss
\hline
CfAN50\df & 5000 &$ [-20.3,-19.0] $& 206 &$ 6.7 \times 10^4 $\ss
CfAN65\df & 6500 &$   < -19.5  $    & 208 &$ 1.6 \times 10^5 $\ss
CfAN80\df & 8000 &$   < -20\ph $    & 207 &$ 3.0 \times 10^5 $\ss
CfAN90\df & 9000 &$   < -20.3  $    & 146 &$ 4.4 \times 10^5 $\ss
\hline
\end{tabular}
\end{center}

\bigskip

\begin{center}
TABLE 2\\
SSRS Samples
\end{center}

\begin{center}
\begin{tabular}{|c c c c c|} \hline \rule[-1.ex]{0mm}{3.6ex}
\makebox[7em]{Sample}&\makebox[6em]{$ z_\lim $ (km/s)}%
&\makebox[8em]{$d_\lim $ (kpc)} &\makebox[5em]{$N_{\rm gal}$}&%
\makebox[6em]{$V_T$ (Mpc$^3$)}\ss\hline\rule[-1.ex]{0mm}{3.6ex}%
SSRS50\df & 5000 &    18.3--33  & 222 &$ 6.7 \times 10^4 $\ss
SSRS65\df & 6500 &$ > 22 \ph $& 283 &$ 1.5 \times 10^5 $\ss
SSRS80\df & 8000 &$ > 29.3   $& 220 &$ 2.9 \times 10^5 $\ss
SSRS90\df & 9000 &$ > 33 \ph $& 203 &$ 4.2 \times 10^5 $\ss
\hline
\end{tabular}
\end{center}

\bigskip
\begin{center}
TABLE 3\\
IRAS Samples
\end{center}

\begin{center}
\begin{tabular}{|c c c c c|} \hline \rule[-1.ex]{0mm}{3.6ex}
\makebox[7em]{Sample}&\makebox[6em]{$ z_\lim $ (km/s)}%
&\makebox[8em]{$z^2 f_{60}$ (Jy Mpc$^2$)} &\makebox[5em]{$N_{\rm gal}$}&%
\makebox[6em]{$V_T$ (Mpc$^3$)}\ss\hline\rule[-1.ex]{0mm}{3.6ex}%
IRAS45\df & 4500 &$ >\ 3920 $& 319 &$ 3.3 \times 10^5 $\ss
IRAS55\df & 5500 &$ >\ 5856 $& 404 &$ 6.1 \times 10^5 $\ss
IRAS65\df & 6500 &$ >\ 8180 $& 359 &$ 1.0 \times 10^6 $\ss
IRAS80\df & 8000 &$ > 12390 $& 314 &$ 1.9 \times 10^6 $\ss
\hline
\end{tabular}
\end{center}

\newpage

\begin{center}
TABLE $4a$\\
CfA Results, Real Space
\end{center}

\begin{center}
\begin{tabular}{|c c c c c c|} \hline \rule[-1.ex]{0mm}{3.6ex}
\makebox[5em]{Sample}&\makebox[5em]{$ \theta z_\lim $ (Mpc)}%
&\makebox[6.5em]{$\gamma$} &\makebox[6em]{$r_0$ (Mpc)}&%
\makebox[6em]{$S_3$}&\makebox[6.8em]{$S_4$}\ss
\hline\rule[-1.ex]{0mm}{3.6ex}%
CfAN50\df& 1--6  &$ 1.88 \pm 0.03 ~(0.05) $&$ 3.68 \pm 0.11 ~(0.21) $
	&$ 1.98 \pm 0.22 ~(0.70) $&$ 6.0 \pm 3.6 ~(4.1) $\ss
CfAN65\df& 1--8  &$ 1.82 \pm 0.02 ~(0.04) $&$ 4.98 \pm 0.10 ~(0.17) $
	&$ 1.87 \pm 0.13 ~(0.52) $&$ 5.6 \pm 4.8 ~(3.1) $\ss
CfAN80\df& 2--10 &$ 1.76 \pm 0.05 ~(0.09) $&$ 7.09 \pm 0.53 ~(0.88) $
	&$ 1.91 \pm 0.18 ~(0.50) $&$ 5.4 \pm 2.2 ~(1.3) $\ss
CfAN90\df& 3--14 &$ 2.26 \pm 0.13 ~(0.12) $&$ 5.75 \pm 0.70 ~(0.64) $
	&$ 2.03 \pm 0.45 ~(0.43) $&$ 9.8 \pm 3.7 ~(3.9) $\ss
Average\df& 1--14 &  $ 1.86 \pm  0.07 ~(0.03)$&$ 4.56 \pm 0.50 ~(0.13)$
	&  $ 1.95 \pm  0.04 ~(0.26)$&$ 6.3 \pm 1.1 ~(1.6)$\ss
\hline
\end{tabular}
\end{center}

\bigskip

\begin{center}
TABLE $4b$\\
CfA Results, Redshift Space
\end{center}

\begin{center}
\begin{tabular}{|c c c c c c|} \hline \rule[-1.ex]{0mm}{3.6ex}
\makebox[5em]{Sample}&\makebox[5em]{$R$ (Mpc)}%
&\makebox[6.5em]{$\gamma$} &\makebox[6em]{$s_0$ (Mpc)}&%
\makebox[6em]{$S_3$}&\makebox[6.8em]{$S_4$}\ss
\hline\rule[-1.ex]{0mm}{3.6ex}%
CfAN50\df& 2--7  &$ 1.62 \pm 0.08 ~(0.10) $&$ 4.47 \pm 0.27 ~(0.33) $
	&$ 1.77\pm 0.13 ~(0.33) $&$ ~4.9 \pm 0.9 ~(1.7) $\ss
CfAN65\df& 3--8  &$ 1.89 \pm 0.23 ~(0.36) $&$ 4.81 \pm 0.86 ~(1.33) $
	&$ 1.57 \pm 0.16 ~(0.53) $&$ ~2.9 \pm 1.0 ~(2.7) $\ss
CfAN80\df& 4--13 &$ 1.56 \pm 0.06 ~(0.10) $&$ 8.13 \pm 0.64 ~(1.06) $
	&$ 2.34 \pm 0.14 ~(0.62) $&$ ~7.8 \pm 1.0 ~(3.5) $\ss
CfAN90\df& 8--16 &$ 1.82 \pm 0.22 ~(0.45) $&$ 6.83 \pm 1.61 ~(3.31) $
	&$ 2.57 \pm 0.29 ~(0.85) $&$ 10.2 \pm 2.4 ~(6.1) $\ss
Average\df& 2--16 &$ 1.61 \pm 0.05 ~(0.07)$&$ 4.81 \pm 0.67 ~(0.31) $
	&$ 1.88 \pm 0.31 ~(0.24)$&$ ~5.1 \pm 1.2 ~(1.3)$\ss
\hline
\end{tabular}
\end{center}

\newpage

\begin{center}
TABLE $5a$\\
SSRS Results, Real Space
\end{center}

\begin{center}
\begin{tabular}{|c c c c c c|} \hline \rule[-1.ex]{0mm}{3.6ex}
\makebox[5em]{Sample}&\makebox[5em]{$ \theta z_\lim $ (Mpc)}%
&\makebox[6.5em]{$\gamma$} &\makebox[6em]{$r_0$ (Mpc)}&%
\makebox[6em]{$S_3$}&\makebox[6.8em]{$S_4$}\ss
\hline\rule[-1.ex]{0mm}{3.6ex}%
SSRS50\df& 1--6  &$ 1.64 \pm 0.02 ~(0.03) $&$ 3.60 \pm 0.33 ~(0.45) $
	&$ 1.72 \pm 0.14 ~(0.30) $&$ 4.8 \pm 4.7 ~(1.6) $\ss
SSRS65\df& 1--8  &$ 1.57 \pm 0.01 ~(0.03) $&$ 5.27 \pm 0.14 ~(0.47) $
	&$ 2.09 \pm 0.15 ~(0.82) $&$ 6.2 \pm 4.6 ~(4.4) $\ss
SSRS80\df& 2--7 &$ 1.64 \pm 0.09 ~(0.10) $&$ 5.69 \pm 0.57 ~(0.69) $
	&$ 1.77 \pm 0.18 ~(0.23) $&$ 5.0 \pm 3.6 ~(1.8) $\ss
SSRS90\df& 2--12 &$ 1.71 \pm 0.04 ~(0.06) $&$ 4.80 \pm 0.24 ~(0.36) $
	&$ 1.78 \pm 0.48 ~(0.93) $&$ 6.3 \pm 2.3 ~(5.7) $\ss
Average\df& 1--12 &$ 1.62 \pm 0.03 ~(0.02) $&$ 4.70 \pm 0.46 ~(0.23) $
	&$ 1.77 \pm 0.05 ~(0.17) $&$ 5.4 \pm 0.4 ~(2.2) $\ss
\hline
\end{tabular}
\end{center}

\bigskip

\begin{center}
TABLE $5b$\\
SSRS Results, Redshift Space
\end{center}

\begin{center}
\begin{tabular}{|c c c c c c|} \hline \rule[-1.ex]{0mm}{3.6ex}
\makebox[5em]{Sample}&\makebox[5em]{$R$ (Mpc)}%
&\makebox[6.5em]{$\gamma$} &\makebox[6em]{$s_0$ (Mpc)}&%
\makebox[6em]{$S_3$}&\makebox[6.8em]{$S_4$}\ss
\hline\rule[-1.ex]{0mm}{3.6ex}%
SSRS50\df& 2--6  &$ 1.32 \pm 0.08 ~(0.11) $&$ 4.79 \pm 0.35 ~(0.51) $
	&$ 1.65 \pm 0.10 ~(0.33) $&$ 4.0 \pm 0.5 ~(1.8) $\ss
SSRS65\df& 3--9  &$ 1.05 \pm 0.02 ~(0.05) $&$ 9.16 \pm 0.32 ~(0.77) $
	&$ 1.97 \pm 0.06 ~(0.48) $&$ 5.5 \pm 0.4 ~(2.2) $\ss
SSRS80\df& 4--10 &$ 1.30 \pm 0.11 ~(0.13) $&$ 9.15 \pm 1.41 ~(1.71) $
	&$ 1.97 \pm 0.11 ~(0.35) $&$ 5.6 \pm 0.8 ~(1.7) $\ss
SSRS90\df& 5--11 &$ 1.25 \pm 0.22 ~(0.25) $&$ 8.15 \pm 2.49 ~(2.85) $
	&$ 1.83 \pm 0.17 ~(0.37) $&$ 5.3 \pm 1.4 ~(2.2) $\ss
Average\df& 2--11 &$ 1.12 \pm 0.08 ~(0.04) $&$ 6.33 \pm 1.36 ~(0.41) $
	&$ 1.83 \pm 0.09 ~(0.19) $&$ 5.1 \pm 0.5 ~(1.0) $\ss
\hline
\end{tabular}
\end{center}

\newpage

\begin{center}
TABLE $6a$\\
IRAS Results, Real Space
\end{center}

\begin{center}
\begin{tabular}{|c c c c c c|} \hline \rule[-1.ex]{0mm}{3.6ex}
\makebox[5em]{Sample}&\makebox[5em]{$ \theta z_\lim $ (Mpc)}%
&\makebox[6.5em]{$\gamma$} &\makebox[6em]{$r_0$ (Mpc)}&%
\makebox[6em]{$S_3$}&\makebox[6.8em]{$S_4$}\ss
\hline\rule[-1.ex]{0mm}{3.6ex}%
IRAS45\df& 2--13 &$ 1.89 \pm 0.02 ~(0.04) $&$ 3.85 \pm 0.15 ~(0.27) $
	&$ 2.06 \pm 0.25 ~(0.38) $&$ ~7.4 \pm 6.1 ~(2.9) $\ss
IRAS55\df& 2--17 &$ 1.75 \pm 0.02 ~(0.03) $&$ 4.04 \pm 0.10 ~(0.18) $
	&$ 2.44 \pm 0.52 ~(0.48) $&$ ~9.4 \pm 5.9 ~(4.0) $\ss
IRAS65\df& 3--20 &$ 1.76 \pm 0.11 ~(0.15) $&$ 3.80 \pm 0.44 ~(0.60) $
	&$ 2.33 \pm 1.35 ~(0.69) $&$ 11.7 \pm 9.6 ~(9.2) $\ss
IRAS80\df& 4--16 &$ 1.87 \pm 0.21 ~(0.32) $&$ 4.00 \pm 1.23 ~(1.14) $
	&$ 3.41 \pm 2.11 ~(1.22) $&$ ~24 \; \pm \; 34 ~(23) $\ss
Average\df& 2--20 &$ 1.80 \pm 0.04 ~(0.02) $&$ 3.97 \pm 0.06 ~(0.14) $
	&$ 2.22 \pm 0.16 ~(0.30) $&$ ~9.2 \pm 1.5 ~(3.9) $\ss %
\hline
\end{tabular}
\end{center}

\bigskip

\begin{center}
TABLE $6b$\\
IRAS Results, Redshift Space
\end{center}

\begin{center}
\begin{tabular}{|c c c c c c|} \hline \rule[-1.ex]{0mm}{3.6ex}
\makebox[5em]{Sample}&\makebox[5em]{$R$ (Mpc)}%
&\makebox[6.5em]{$\gamma$} &\makebox[6em]{$s_0$ (Mpc)}&%
\makebox[6em]{$S_3$}&\makebox[6.8em]{$S_4$}\ss
\hline\rule[-1.ex]{0mm}{3.6ex}%
IRAS45\df& 3--9 &$ 1.36 \pm 0.23 ~(0.23) $&$ 5.37 \pm 1.23 ~(1.20) $
	&$ 2.28 \pm 0.41 ~(0.25) $&$ 8.8 \pm 3.4 ~(2.2) $\ss
IRAS55\df& 4--10 &$ 1.50 \pm 0.17 ~(0.15) $&$ 4.65 \pm 0.90 ~(0.79) $
	&$ 1.93 \pm 0.21 ~(0.33) $&$ 6.6 \pm 1.5 ~(2.8) $\ss
Average\df& 3--10 &$ 1.44 \pm 0.09 ~(0.14) $&$ 4.90 \pm 0.45 ~(0.73) $
	&$ 2.07 \pm 0.23 ~(0.26)$&$ 7.5 \pm 1.4 ~(2.1) $\ss
\hline
\end{tabular}
\end{center}

\newpage

\begin{center}
TABLE $7$\\
Results for $\Omega^{4/7}/b$
\end{center}

\begin{center}
\begin{tabular}{|c c c c c|} \hline \rule[-1.ex]{0mm}{3.6ex}
\makebox[7.5em]{Sample}&\makebox[6em]{$r_\max$ (Mpc)}%
&\makebox[6em]{$\xi_2(r_\max)$}&\makebox[6em]{$\Omega^{4/7}/b ~ (r_0)$}%
&\makebox[6em]{$\Omega^{4/7}/b~ (r_\max)$}\ss
\hline\rule[-1.ex]{0mm}{3.6ex}%
CfA50\df& 5.18 & 0.53 &$  ~0.48 \pm 0.24  $&$  ~0.63 \pm 0.26  $\ss
CFA65\df& 5.52 & 0.83 &$  -0.10 \pm 0.79  $&$  -0.16 \pm 0.78  $\ss
CfA80\df& 9.75 & 0.57 &$  ~0.33 \pm 0.44  $&$  ~0.43 \pm 0.47  $\ss
CfA90\df& 11.3 & 0.22 &$  ~0.48 \pm 1.45  $&$  ~0.98 \pm 1.61  $\ss
CfA Average\df& --- & --- &$ 0.41 \pm 0.10 $&$ ~0.53 \pm 0.15 $\ss
\hline
SSRS50\df& 4.73 & 0.64 &$ ~0.58 \pm 0.35  $&$ ~0.73 \pm 0.41 $\ss
SSRS65\df& 7.44 & 0.58 &$ ~0.92 \pm 0.23  $&$ ~1.24 \pm 0.31 $\ss
SSRS80\df& 7.91 & 0.58 &$ ~0.99 \pm 0.52  $&$ ~1.19 \pm 0.57 $\ss
SSRS90\df& 8.79 & 0.36 &$ ~1.07 \pm 0.82  $&$ ~1.59 \pm 0.90 $\ss
SSRS Average\df& --- & --- &$ ~0.85 \pm 0.10 $&$ ~1.10 \pm 0.16 $\ss
\hline
IRAS45\df& 7.04 & 0.32 &$ ~0.71 \pm 0.56  $&$ ~1.27 \pm 0.64 $\ss
IRAS55\df& 7.60 & 0.33 &$ ~0.32 \pm 0.46  $&$ ~0.57 \pm 0.50 $\ss
IRAS Average\df& --- & --- &$ ~0.48 \pm 0.25 $&$  ~0.84 \pm 0.45 $\ss
\hline
\end{tabular}
\end{center}

\vfill\eject

\def\hf{\hfill}

\begin{center}
Table 8\\
Estimates for $S_3$ and $S_4$
\end{center}

\begin{center}
\begin{tabular}{|c l c c|} \hline \rule[-0.6ex]{0mm}{3.2ex}
\makebox[11em]{Reference} & \makebox[6em]{Sample} &
\makebox[6em]{$S_3$ or $3 Q_3$} &\makebox[5em]{$S_4$ or $16 Q_4$}
\ss \hline
Groth \& Peebles 1977\df&Lick-Zwicky &$ 3.9 \pm 0.6 $& --- \ss
Fry \& Peebles 1978\df&Lick-Zwicky & --- &$ 46 \pm ~8 $\ss
Szapudi et al. 1992\df&Lick &$ 4.3 \pm 0.2 $&$ 31 \pm ~5 $\ss
Peebles 1980\df&CfA&$ 2.4 \pm 0.2 $& --- \ss
This paper\df&CfA&$ 2.0 \pm 0.3 $&$ 6.3 \pm 1.6 $\ss
This paper\df&SSRS&$ 1.8 \pm 0.2 $&$ 5.4 \pm 2.2 $\ss
Meiksin et al. 1992\df&IRAS &$ 2.2 \pm 0.2 $&$ 10 \pm ~3 $\ss
This paper\df&IRAS &$ 2.2 \pm 0.3 $&$ 9.2 \pm 3.9 $\ss
Gazta\~naga 1992\df &CfA\hf(redshift)&$ 1.9 \pm 0.1 $&$ 4.1 \pm 0.6$\ss
This paper\df&CfA\hf(redshift)&$ 1.9 \pm 0.2 $&$ 5.1 \pm 1.3 $\ss
Gazta\~naga 1992\df&SSRS\hf~(redshift)&$ 2.0 \pm 0.1 $&$ 5.0 \pm 0.9$\ss
This paper\df& SSRS\hf~(redshift)&$1.8 \pm 0.2 $&$ 5.2 \pm 1.3 $\ss
Bouchet \etal 1993\df&IRAS\hf~(redshift)&$ 1.5 \pm 0.5$&$4.4 \pm 3.7$\ss
This paper\df&IRAS\hf~(redshift)&$ 2.1 \pm 0.3 $&$ 7.5 \pm 2.1 $\ss
\hline
\end{tabular}
\end{center}

\newpage

\section{Figure Captions}

\bigskip

\noindent
Fig.~1.---Average correlation function, $\xibar_2$ for each sample
in the CfA ($a$), SSRS ($b$) and IRAS ($c$) catalogue.
Open squares correspond to $\xibar_2$ for spherical (redshift)
cells as a function of the radius $R$ of the cell.
Filled triangles correspond to $\xibar_2$ for conical (real space)
cells as a function of the radius $\theta z$ at the base of the cone.
The dashed lines are the best power-law fit weighted by the errors.

\bigskip

\noindent
Fig.~2.---Values of $\xibar_3$ (triangles) and  $\xibar_4$ (squares),
as a function of $\xibar_2$ for spherical (redshift) cells.
There is one graph for each sample in the
CfA ($a$), SSRS ($b$) and IRAS ($c$) catalogue.
The dashed lines are the hierarchical law: $\xibar_J=S_J \xibar_2^{J-1}$
where  $S_3$ and $S_4$ are fitted with the data weighted by the errors.

\bigskip

\noindent
Fig.~3.---Values of $\xibar_3$ (triangles) and  $\xibar_4$ (squares), as a
function of $\xibar_2$ for conical (real space) cells.
There is one graph for each sample in the CfA ($a$), SSRS ($b$),
and IRAS ($c$) catalogue.
The dashed lines are the hierarchical law: $\xibar_J=S_J \xibar_2^{J-1}$
where  $S_3$ and $S_4$ are fitted with the data weighted by the errors.

\end{document}